\documentclass[letterpaper]{article}

\usepackage[T1]{fontenc}
\usepackage{xr}
\usepackage{amssymb}
\usepackage{geometry}
\usepackage{setspace}
\usepackage{amsmath}
\usepackage{siunitx}
\usepackage{subcaption}

\usepackage[style = chem-acs, articletitle=true]{biblatex}
\usepackage{graphicx}
\usepackage{float}
\newfloat{scheme}{htbp}{los}
\floatname{scheme}{Scheme}
\floatname{chart}{Chart}
\newfloat{graph}{htbp}{loh}

\usepackage{chemformula} % Formulas using \ch{}
\usepackage[version = 4]{mhchem} % Formulas using \ce{}
\usepackage{amssymb}
\newcommand{\beginsupplement}{%
	        \setcounter{table}{0}
	        \renewcommand{\thetable}{S\arabic{table}}%
	        \setcounter{figure}{0}
	        \renewcommand{\thefigure}{S\arabic{figure}}%
	        \renewcommand{\thesection}{S\arabic{section}}%
	        \renewcommand{\theequation}{S\arabic{equation}}%
	     }
         
\usepackage{authblk}
\usepackage{comment} %only for comment package
\usepackage{lipsum}
\usepackage{xcolor}

\author[1]{Ivan Villani}
\author[2,3]{Samuele Fracassi}
\author[2,3]{Niccolò Traverso Ziani}
\author[2,3]{Maura Sassetti}
\author[3]{Matteo Carrega$^{*}$}
\author[4]{Vaidotas Mišeikis}
\author[4]{Camilla Coletti}
\author[1]{Fabio Beltram}
\author[5]{Kenji Watanabe}
\author[6]{Takashi Taniguchi}
\author[1]{Sergio Pezzini$^{*}$}
\author[1]{Stefan Heun$^{*}$}

\affil[1]{Istituto Nanoscienze-CNR, NEST-Scuola Normale Superiore, Piazza San Silvestro 12, 56127 Pisa, Italy}
\affil[2]{Dipartimento di Fisica, Università di Genova, Genova, Italy}
\affil[3]{CNR-SPIN, Via Dodecaneso 33, 16146 Genova, Italy}
\affil[4]{Center for Nanotechnology Innovation, Laboratorio NEST, Istituto Italiano di Tecnologia, 56127 Pisa, Italy}
\affil[5]{Research Center for Electronic and Optical Materials, National Institute for Materials Science, 1-1 Namiki, Tsukuba 305-0044, Japan}
\affil[6]{Research Center for Materials Nanoarchitectonics, National Institute for Materials Science, 1-1 Namiki, Tsukuba 305-0044, Japan}

\title{Coherent Electronic Lensing in Proximitized Graphene}

\date{* matteo.carrega@cnr.it and sergio.pezzini@cnr.it and stefan.heun@cnr.it}

\begin{document}

\maketitle

\begin{abstract} % 170 words
Spatially resolved control of coherent electron transport offers a route to understanding and engineering interference phenomena in hybrid superconducting devices. Here, we use Scanning Gate Microscopy (SGM) to investigate Fabry-Pérot (FP) interference in the n-p-n cavity of a ballistic graphene Josephson junction. Spatial maps of the superconducting switching current and normal-state conductance reveal circular interference fringes arising from two distinct effects of the SGM tip potential. A global shift of the Fermi wave vector modulates the FP resonances of the cavity, while a local perturbation produces additional oscillations that persist at fixed Fermi wave vector as the tip bias is varied. An analytical model identifies the origin of this local response with coherent electron lensing, in which the tip modifies the phase accumulated along interfering trajectories. The measured fringe patterns in the switching current indicate that this mechanism affects superconducting transport, consistent with tip-induced shifts of the resonance conditions for Andreev bound states. These results establish SGM as a tunable local lens for controlling interference in superconducting ballistic graphene devices.
\end{abstract}

%\section*{Keywords}
%Graphene, Josephson junction, Fabry-Pérot, Scanning Gate Microscopy

%\section{Introduction}

Electron quantum optics provides a powerful framework for exploring coherent transport in mesoscopic conductors \cite{Chen2016,  Chakraborti2024}, in which the propagation of electrons can be understood in terms of trajectories, refraction, focusing, and interference. The solid-state electronic counterpart of optical phenomena has been introduced in high mobility two-dimensional electron gases (2DEGs), demonstrating coherent electronic beam control, electronic interferometry, and collider-like experiments \cite{Ji2003, Bocquillon2014, Acciai2019,  Bartolomei2020, Nakamura2020, Carrega2021, Ronetti2024}.

In this context, graphene is regarded as a key experimental platform, thanks to the linear dispersion relation and the chiral nature of the charge carriers that give rise to transport phenomena without direct analogues in conventional semiconductor systems \cite{Novoselov2005,Zhang2005}. For instance, the possibility of electrostatically defining p-n junctions, which act as interfaces between electron-like and hole-like states, allows for unusual electron-optical functionalities, due to Klein tunneling and negative refraction \cite{Katsnelson2006, Allain2011, Lee2015}, leading to carrier guiding, focusing, and Veselago lensing \cite{Cheianov2007, Williams2011, Zhang2022}.

An important manifestation of phase-coherent electron propagation is Fabry-Pérot (FP) interferometry \cite{Carrega2021, Deprez2021, Kim2024}, in which carriers undergo multiple reflections between partially transmitting interfaces, producing conductance oscillations as a function of carrier density, gate voltage, and magnetic field. In graphene, such interference can arise in electrostatically defined n-p-n junctions, where the two interfaces form a cavity for charge carriers \cite{Shytov2008}. There, the interplay between Klein tunneling, which favors transmission at near-normal incidence, and the angle-dependent reflection at the interfaces gives rise to an interference pattern that reflects the chiral nature of graphene quasiparticles \cite{Young2009, Nam2011}.

Although conductance measurements reveal the presence of coherent interference, they generally provide only indirect information about the spatial trajectories responsible for it. Scanning Gate Microscopy (SGM) can overcome this limitation. Indeed, by using a biased, movable tip to locally perturb the electrostatic potential of a device, local maps of conductance can be recorded. In conventional 2DEGs, SGM has proven useful to image coherent electron flow, to enable local carrier manipulation, and to build interferometric setups \cite{Topinka2001, Crook2003, Cunha2006, Aidala2007, Martins2007, Paradiso2012, Toussaint2018}. In graphene, previous work has used SGM to investigate coherent transport, demonstrating guided electronic states, disorder-induced phenomena, and electron-optical phenomena \cite{Connolly2011, Connolly2012,  Garcia2013, Herbschleb2015, Bours2017, Cabosart2017, Brun2019, Marguerite2019, Brun2021, Gold2021}.

Introducing superconducting contacts adds a further dimension to this picture. Graphene Josephson junctions support phase-coherent transport, in which supercurrent is mediated by Andreev processes: an electron incident on a superconducting interface is reflected as a hole, while a Cooper pair enters the superconductor. The resulting electron–hole dynamics provides a route to studying superconducting proximity effects in a material whose carrier density and polarity remain electrostatically tunable. Experiments on graphene Josephson junctions have demonstrated ballistic supercurrent transport, quantum oscillations of the critical current, and FP interference in the superconducting state \cite{Calado2015, BenShalom2016, Allen2017}. FP oscillations have been shown to modulate the supercurrent, with local maxima (minima) in the switching current aligned with local maxima (minima) in the conductance \cite{Calado2015, BenShalom2016, Allen2017, Villani2025}. However, how the spatial structure of coherent transport evolves when normal-state electron optics is coupled to superconducting correlations is still an open question.

In this context, spatially resolved measurements are particularly valuable because they can provide information that is not readily accessible from global transport characteristics alone \cite{Kaperek2022, Maji2024, maji2026arxiv}. Recently, SGM has been successfully applied to imaging defects in superconducting resonators \cite{Hagedus2025} and Andreev reflections in graphene \cite{Bhandari2020}, as well as modulating supercurrent flow in Josephson junctions \cite{Lombardi2025, villani_2026_SGM}.

Here, we manipulate Fabry-Pérot interference in a ballistic graphene Josephson junction employing SGM. Differently from previous SGM experiments on graphene-based devices, where the tip was used to create an electrostatic cavity \cite{Brun2019,  Brun2021}, in our work the SGM tip is used to actively control the properties of the n-p-n cavity that already exists in the sample. We experimentally observe a series of spatial circular fringes in the switching current, correlating with analogous patterns in the normal-state conductance. By combining bias and gate-dependent SGM measurements with analytical modeling, we reveal that the fringes are generated by two effects. First, a global tip-induced renormalization of the Fermi wave vector shifts the FP oscillations. Second, the local nature of the tip potential produces oscillations at fixed Fermi wave vector, arising from a coherent lensing effect. Interestingly, the same mechanism accounts also for the fringe pattern observed in the superconducting regime, where Andreev bound states and the Fabry-Pérot cavity coexist. The tip locally modifies the quantum phase of interfering trajectories and thus allows to control quantum-coherent transport across hundreds of nanometers both in the normal and superconducting state.

%\section{Results and Discussion}

\begin{figure}[!tbh] % Fig.1, caption 247 words
	\centering
	\includegraphics[width=\linewidth]{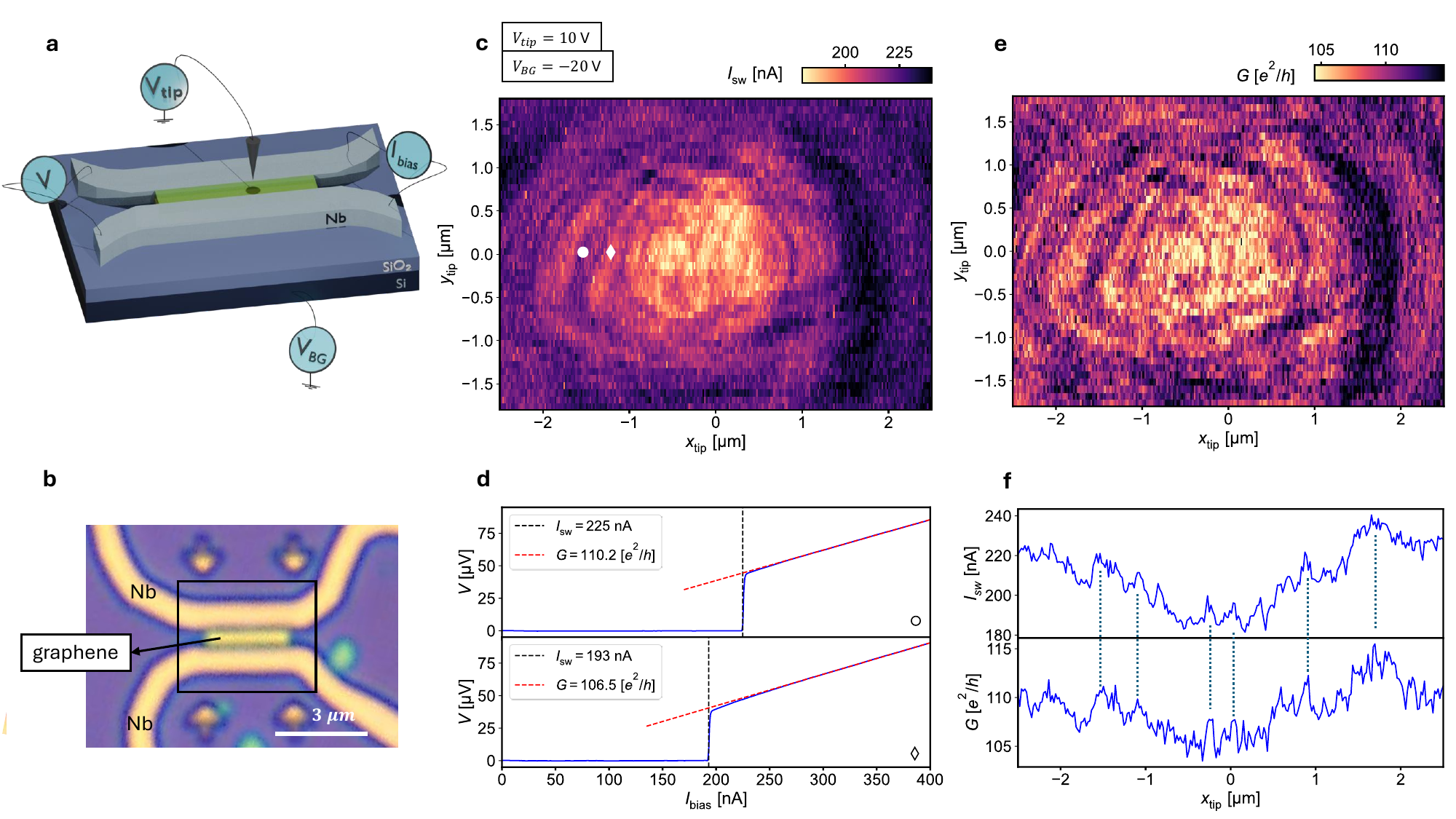}
	\caption{(\label{Fig1} \textbf{a}) Three-dimensional schematics of the graphene Josephson junction and SGM setup. Graphene is encapsulated in two hBN flakes and edge-contacted by Nb superconducting leads. The biased SGM tip scans above the device at fixed distance. (\textbf{b}) Optical microscopy image of the investigated device (junction length $L=600$~nm and width $W=3\;\mu$m). The black rectangle indicates the scanning area of the maps shown in panels \textbf{c} and \textbf{e}. (\textbf{c}) SGM map of switching current $I_{\mathrm{sw}}$ as a function of the SGM tip coordinates. Tip bias $V_{\mathrm{tip}}=10$~V, back-gate voltage $V_{\mathrm{BG}}=-20$~V. The white dot and diamond indicate the position of the $V$-$I_{\mathrm{bias}}$ curves shown in panel \textbf{d}. (\textbf{d}) Representative $V$-$I_{\mathrm{bias}}$ curves from map in panel \textbf{c} at $y_{\mathrm{tip}}=0$ and different $x_{\mathrm{tip}}$ coordinates, showing the tip-induced modulation of $I_{\mathrm{sw}}$ and $G$. (\textit{Upper panel}) $x_{\mathrm{\mathrm{tip}}}=-1.54\;\mu$m, white dot in Fig.~\ref{Fig1}c. (\textit{Lower panel}) $x_{\mathrm{\mathrm{tip}}}=-1.21\;\mu$m, white diamond in Fig.~\ref{Fig1}c. $I_{\mathrm{sw}}$ (black vertical dashed lines) is obtained by setting a threshold on the measured voltage drop $V$, while $G$ is obtained from a best-fit in the dissipative branch of the $V$-$I_{\mathrm{bias}}$ curves (red dashed lines). (\textbf{e}) SGM map of junction conductance $G$ obtained from the same dataset of panel \textbf{c}, by fitting the dissipative branch of $V$-$I_{\mathrm{bias}}$ curves. (\textbf{f}) Representative line-cuts of $I_{\mathrm{sw}}$ (upper panel) and of $G$ (lower panel) from panels \textbf{c} and \textbf{e}, obtained as the average of three lines around $y_{\mathrm{\mathrm{tip}}}=0$, showing that $I_{\mathrm{sw}}$ and $G$ share the same oscillatory pattern. Dotted vertical lines highlight the alignment between local maxima and minima in $I_{\mathrm{sw}}$ and $G$.
    }
\end{figure}

We investigate a ballistic graphene Josephson junction fabricated with Niobium (\ce{Nb}) superconducting edge-contacts. We employ high quality graphene crystals grown by Chemical Vapor Deposition (CVD) \cite{Miseikis2015,Pezzini2020} and encapsulated in 30 nm-thick hexagonal Boron Nitride (hBN) flakes. The hBN-graphene-hBN heterostructure is placed on a \ce{Si/SiO_2} substrate that acts as a back-gate via the application of a voltage $V_{\mathrm{BG}}$. Fabrication of the device follows the procedure described in Refs.~\cite{Villani2025,villani2026diode_arxiv,villani_2026_SGM} and briefly summarized in the Supporting Information. A three-dimensional sketch of the Josephson junction and of the measurement setup is reported in Fig.~\ref{Fig1}a. An optical microscopy image of the device is shown in Fig.~\ref{Fig1}b.

All measurements presented in this work were performed in a Janis cryostat equipped with an Attocube AFM head operating at $T=300$~mK. Electrical lines for sample measurement and to AFM components are equipped with cryogenic filters to preserve the superconducting properties of the investigated device. The SGM maps reported here are obtained by working in \textit{flying mode}: the tip is scanned above the sample in non-contact mode at a constant tip-to-graphene distance of $d=100$~nm (corresponding to a tip-to-sample distance $d=70$~nm, considering the thickness of $30$~nm of the top hBN flake). By applying a voltage bias $V_{\mathrm{tip}}$, the tip acts as a movable local gate.

In Fig.~\ref{Fig1}c we report the switching current $I_{\mathrm{sw}}$ of the device as a function of tip position $(x_{\mathrm{\mathrm{tip}}},y_{\mathrm{\mathrm{tip}}})$, acquired by setting the back-gate voltage to $V_{\mathrm{BG}}=-20$~V (corresponding to a hole carrier density $n=8.2\times 10^{11}$~cm$^{-2}$), and applying a positive tip voltage bias $V_{\mathrm{tip}}=10$~V. The map was obtained by measuring $V$-$I_{\mathrm{bias}}$ curves at each pixel of a grid defined in the $5\times 3.6\;\mu$m$^2$ area indicated by the black rectangle in Fig.~\ref{Fig1}b. Each $V$-$I_{\mathrm{bias}}$ curve was obtained by sweeping the current bias $I_{\mathrm{bias}}$ from $0$ to $400$~nA, and the switching current $I_{\mathrm{sw}}$ was extracted by setting a threshold on the measured voltage drop $V$. 

While scanning over the junction, the biased SGM tip modulates $I_{\mathrm{sw}}$. {Thanks to the high resolution of} the SGM map of Fig.~\ref{Fig1}c, two contributions in the tip-induced modulation of $I_{\mathrm{sw}}$ can be distinguished. First, a reduction of $I_{\mathrm{sw}}$ is observed, which is maximum when the tip is moved to the junction center. Similar behavior has been recently reported in Refs.~\cite{Lombardi2025,villani_2026_SGM}. The tip creates a potential modulation that is analogous to that of an additional top gate with a different (smaller, in the experimental conditions employed here) lever arm compared to the back-gate. In the hole doping regime and for positive $V_{\mathrm{tip}}$, a reduction of the supercurrent is observed in Fig.~\ref{Fig1}c. In the following we refer to this effect as the background modulation of $I_{\mathrm{sw}}$.

Interestingly, in addition to the background modulation described above, we also observe multiple circular fringes that intersect each other. This additional tip-induced effect on $I_{\mathrm{sw}}$ can be clearly appreciated in the representative $V$-$I_{\mathrm{bias}}$ curves shown in Fig.~\ref{Fig1}d. These curves are taken at $y_{\mathrm{\mathrm{tip}}}=0$, at different $x_{\mathrm{\mathrm{tip}}}$-coordinates: on a fringe maximum ($x_{\mathrm{\mathrm{tip}}}=-1.54\;\mu$m, white dot in Fig.~\ref{Fig1}c), giving $I_{\mathrm{sw}} = 225$~nA (top panel), and on a fringe minimum ($x_{\mathrm{\mathrm{tip}}}=-1.21\;\mu$m, white diamond in Fig.~\ref{Fig1}c), yielding $I_{\mathrm{sw}} = 193$~nA (bottom panel). A difference of $\Delta I_{\mathrm{sw}} = 32$~nA is observed between the two curves. An additional $I_{\mathrm{sw}}$ map obtained with opposite tip bias as in Fig.~\ref{Fig1}c shows comparable features and is presented in the Supporting Information.

To further investigate these spatial features, it is useful to relate maps of $I_{\mathrm{sw}}$ to the corresponding ones in the dissipative regime, i.e., at $I_{\mathrm{bias}}>I_{\mathrm{sw}}$. Indeed, from the the slope of measured $V$-$I_{\mathrm{bias}}$ curves it is also possible to extract the sample resistance $R$, obtaining the conductance as $G = 1/R$. In Fig.~\ref{Fig1}d, linear fits in the current interval $300-400$~nA are shown as red dashed lines. A larger $G=110.2\pm0.1\;e^2/h$ is obtained from the $V$-$I_{\mathrm{bias}}$ curve corresponding to the larger $I_{\mathrm{sw}}$ (top panel), as compared to $G=106.5\pm0.1\;e^2/h$ obtained from the second $V$-$I_{\mathrm{bias}}$ curve (bottom panel). By repeating the same process for all curves, we obtain the SGM map of the conductance $G$ shown in Fig.~\ref{Fig1}e. In this map we observe a clear spatial modulation analogous to that pertaining to $I_{\mathrm{sw}}$. Multiple circular intersecting fringes are present, which share the same spatial pattern as in the $I_{\mathrm{sw}}$ map. A global background modulation is also observed, with minimum conductance with the tip above the junction center.

It is well known that in Josephson junctions a positive correlation exists between normal-state resistance $R_n$ and switching current $I_{sw}$. While both quantities are gate-tunable, the $I_{sw} \cdot R_n$ product is observed to be constant and independent of back-gate voltage \cite{Calado2015, BenShalom2016}. This direct link is also observed here in the switching current reported in Fig.~\ref{Fig1}c and in the conductance map of Fig.~\ref{Fig1}e. Here we take the measured $R = 1/G$ as an approximation for $R_n$, noting that this is an underestimation which neglects the excess current (indeed, in the chosen parameter range, $V$ remains much smaller than the gap of the superconductor, resulting in an enhancement of $G$ with respect to the normal state \cite{tinkham1996}). This direct link is further clarified in Fig.~\ref{Fig1}f, where we compare representative line-cuts of $I_{\mathrm{sw}}$ and $G$. Local maxima and minima in these two plots are aligned, as exemplified by the guiding black dashed lines. For this reason, it is equivalent to consider either the sample conductance or the switching current when investigating the origin of the fringes observed in Figs.~\ref{Fig1}c,e. In the following we focus on the normal-state conductance $G$, measured with a lock-in amplifier applying a $100$~nA AC bias on top of a fixed DC bias of $2\;\mu$A and measuring the voltage drop. Compared to $V$-$I_{\mathrm{bias}}$ curves, this technique allows not only for much faster measurements, but these are also characterized by a lower level of noise. In the Supporting Information we demonstrate that measurements performed either by extracting the sample conductance from $V$-$I_{\mathrm{bias}}$ curves (as shown in Fig.~\ref{Fig1}e) or with a lock-in amplifier give consistent results and share the same oscillatory pattern.

\begin{figure}[!tbh] % Fig.2, caption 322 words
	\centering
    \includegraphics[width=\textwidth]{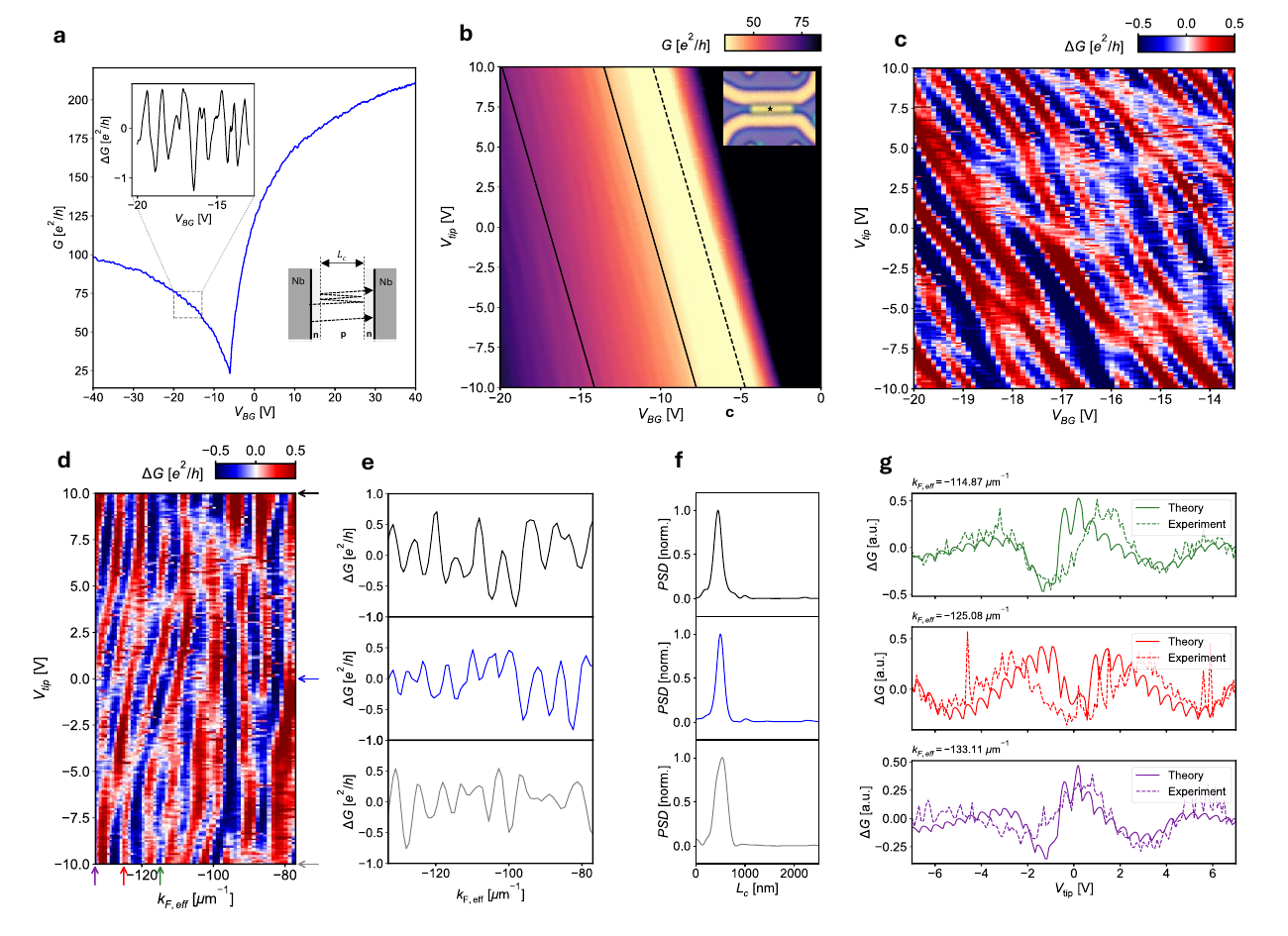}
	\caption{\label{Fig2} (\textbf{a}) Back-gate sweep (conductance $G$ as a function of back-gate voltage $V_{\mathrm{BG}}$). The charge neutrality point (CNP) is shifted to negative back-gate voltages ($V_{\mathrm{BG}}^{\mathrm{CNP}}=-6.3$~V in this specific device). (\textit{Upper inset}) FP oscillations: zoom-in of junction conductance variation $\Delta G$ as a function of $V_{\mathrm{BG}}$ in the range $(-20,-13)$~V, obtained by subtracting a polynomial background from the conductance $G$ of the main panel. (\textit{Lower inset}) Pictorial sketch of FP interference in the n-p-n cavity. $L_c$ is the length of the p-doped region. (\textbf{b}) Map of conductance $G$ as a function of back-gate voltage $V_{\mathrm{BG}}$ and tip bias $V_{\mathrm{tip}}$. Tip is positioned at junction center, as indicated by the black star in the inset. The black dashed line represents the fitted position of the effective charge neutrality point. The black parallelogram corresponds to the data shown in panel \textbf{d}. The conductance $G$ is measured with a lock-in amplifier by applying a bias large enough to suppress any contribution from the excess current. (\textbf{c}) FP oscillations in the hole doping regime as a function of $V_{\mathrm{BG}}$ and $V_{\mathrm{tip}}$: conductance variation $\Delta G$ obtained from panel \textbf{b} by subtracting a two-dimensional polynomial background. (\textbf{d}) Conductance variation $\Delta G$ as a function of effective Fermi wave vector $k_{\text{F,eff}}$ (see main text) and $V_{\mathrm{tip}}$. The displayed area corresponds to the black parallelogram in panel \textbf{b}. (\textbf{e}) Line-cuts of $\Delta G$ as a function of $k_{\text{F,eff}}$ and (\textbf{f}) corresponding Fourier transform for $V_{\mathrm{tip}}=10$~V (black), $V_{\mathrm{tip}}=0$~V (blue), $V_{\mathrm{tip}}=-10$~V (grey). Position of line-cuts is also indicated by horizontal colored arrows in panel \textbf{d}. (\textbf{g}) Comparison between the analytical model of Eq.~\ref{eq1} and experimental line-cuts as a function of $V_{\mathrm{tip}}$ extracted at fixed $k_{\text{F,eff}}$ from panel \textbf{d} (as indicated  by vertical colored arrows in panel \textbf{d}). The same normalization has been applied to all curves, and a background has been subtracted from $G$ to isolate $\Delta G$, as in panels \textbf{c},\textbf{d}. The model parameters used are $\eta=0.1~\text{V}^{-1}$, $L_c = 530~\text{nm}$, $W=3~\mu\text{m}$, $R=430~\text{nm}$, $y_{\mathrm{tip}}=0$.
    }
\end{figure}

To understand the nature of the observed fringe pattern it is instructive to consider first the conductance behavior in absence of the tip. In Fig.~\ref{Fig2}a we show a back-gate sweep (junction conductance $G$ as a function of back-gate voltage $V_{\mathrm{BG}}$) measured without the tip. As reported for previous examples of edge-contacted graphene Josephson junctions \cite{Calado2015,BenShalom2016,Borzenets2016,Villani2025}, this curve is clearly asymmetric, with a lower conductance on the left side of the charge neutrality point CNP (hole doping) compared to the right side (electron doping). This asymmetry originates from the interplay between Fermi level pinning at graphene-Nb interfaces and doping induced by the back-gate. The graphene region in proximity to the Nb contacts is always electron-doped regardless of the applied $V_{\mathrm{BG}}$ due to Fermi level pinning, while the doping in the central region of the junction is controlled by the back-gate. As a result, an n-p-n cavity forms in the device for p-type doping in the central part of the junction. 
This electrostatic cavity behaves as a FP interferometer, with its interfaces acting as partially reflecting mirrors for the massless Dirac carriers, leading to interference among the internally reflected trajectories \cite{Young2009} (as sketched in the lower inset of Fig.~\ref{Fig2}a). By varying the depth of the cavity, its resonance condition changes, leading to an alternation between constructive and destructive interference, which results in conductance oscillations versus back-gate. The upper inset of Fig.~\ref{Fig2}a highlights the presence of this quantum coherent phenomenon in our device.

We now discuss the dependence of the conductance on $V_{\mathrm{tip}}$, starting by keeping the tip position fixed. In Fig.~\ref{Fig2}b we report the sample conductance $G$ as a function of back-gate voltage $V_{\mathrm{BG}}$ and tip voltage $V_{\mathrm{tip}}$, obtained by sweeping $V_{\mathrm{BG}}$ from $-20$~V to $0$~V at different values of $V_{\mathrm{tip}}$ ranging from $+10$~V to $-10$~V. The tip position is kept fixed above the junction center at a vertical distance $d=100$~nm from the graphene channel (as indicated in the inset to Fig.~\ref{Fig2}b). In analogy with the CNP in the back-gate sweep in Fig.~\ref{Fig2}a, the conductance minimum in each line at finite $V_{\mathrm{tip}}$ can be identified as an effective CNP. The effective CNP shifts as a function of $V_{\mathrm{tip}}$, with a linear dependence indicated by the black dashed line in Fig.~\ref{Fig2}b (obtained by fitting points of minimum conductance in each line at fixed $V_{\mathrm{tip}}$). This linear shift can be attributed to a global effective shift of the chemical potential, in which the tip, when positioned above the junction center, acts as a top-gate with a lever arm of $\sim29\%$ relative to the back-gate.

In addition, a series of inclined oscillating features are visible to the left of the effective CNP. This phenomenology can be better seen in Fig.~\ref{Fig2}c, where we report the detrended conductance $\Delta G$ in the hole doping regime for $-20\text{ V}\le V_{\mathrm{BG}}\le -13.5\text{ V}$, obtained from the data in Fig.~\ref{Fig2}b by removing a polynomial background (see details in the Supporting Information). Clearly, the sample conductance oscillates as both back-gate voltage and tip bias are changed. The line at $V_{\mathrm{tip}}=0$~V nominally corresponds to the back-gate sweep shown in Fig.~\ref{Fig2}a (top inset): comparable amplitude and oscillation period are observed. By applying a finite $V_{\mathrm{tip}}$, the oscillating pattern does not change abruptly, but smoothly evolves.
By inspecting Fig.~\ref{Fig2}c it becomes clear that the tip action cannot be simply reduced to that of a global effective top-gate. Indeed, varying $V_{\mathrm{BG}}$ and $V_{\mathrm{tip}}$ along a diagonal parallel to the effective CNP line leaves the effective chemical potential unchanged. Hence, if the tip could be mapped into a global effective top-gate, features in Fig.~\ref{Fig2}c would appear as a series of inclined but continuous stripes parallel to the effective CNP line, without any crossings (see for example Refs.~\cite{Young2009, Nam2011} where multiple gates are used to investigate FP oscillations). This is not the case here, as maxima and minima exhibit multiple crossings along diagonal lines of constant effective chemical potential, with maxima evolving into minima and vice-versa. This additional behavior suggests that the local nature of the tip-induced perturbation plays a crucial role and has to be taken into proper consideration.

As a first hypothesis, one could think that the described pattern originates from a modification of the cavity length $L_c$ induced by the tip potential. To test this hypothesis, we extracted the cavity length as a function of $V_{\mathrm{tip}}$.
To account for the global contribution of the tip discussed above, in Fig.~\ref{Fig2}d we  plot the $\Delta G$ data as a function of $k_{\text{F,eff}}$. The effective Fermi wave vector $k_{\text{F,eff}}$ is calculated relative to the position of the effective CNP (black dashed line in Fig.~\ref{Fig2}b), $k_{\text{F,eff}}=\sqrt{\pi \alpha (V_{\mathrm{BG}}-V_{\mathrm{CNP,eff}})}$, where $\alpha$ is the back-gate lever arm.
Data in Fig.~\ref{Fig2}d correspond to a parallelogram in Fig.~\ref{Fig2}b with lateral edges parallel to the effective CNP position (indicated in Fig.~\ref{Fig2}b by black solid lines). Representative lines of $\Delta G$ vs $k_{\text{F,eff}}$ are shown in Fig.~\ref{Fig2}e for different values of $V_{\mathrm{tip}}$ ($V_{\mathrm{tip}}=10,0,-10$~V respectively in the top, middle, and bottom panels). An oscillating behavior similar to that observed on the hole doping side of the back-gate sweep (inset to Fig.~\ref{Fig2}a) is observed, showing no relevant variations in terms of oscillation amplitude or frequency for different values of $V_{\mathrm{tip}}$. We extract the cavity length for all values of $V_{\mathrm{tip}}$ by means of Fourier transform (representative FFT spectra corresponding to line-cuts in Fig.~\ref{Fig2}e are reported in Fig.~\ref{Fig2}f) and obtain $L_c = 460 \pm 25$~nm. This value shows little to no variation as $V_{\mathrm{tip}}$ is changed over a wide range of $20$~V, and is compatible with the cavity length value extracted from the back-gate sweep of Fig.~\ref{Fig2}a, $L_c=427\pm 30$~nm. This observation rules out the renormalization of $L_c$ as an explanation for the tip-induced modulation of the conductance at fixed $k_{\text{F,eff}}$. The only remaining interpretation is an action of the tip on the coherent FP interference of ballistic trajectories within the cavity.

To understand the interplay between the tip and the cavity, we developed an analytical approach that generalizes, by taking the tip into account, the one that demonstrates the presence of FP oscillations in a graphene p-n-p junction employed by Shytov et al.~in Ref.~\cite{Shytov2008}. In the absence of the tip, the two-dimensional scattering problem reduces to a one-dimensional one, since, assuming translational invariance along the transverse coordinate $x$ %(see Fig. \ref{Fig1}), 
the scattering states can be labeled by a well-defined transverse momentum component $k_x$. To calculate the transmission probability in the effective 1D system, Shytov et al.~sum the amplitudes of all possible FP interference paths, obtaining a $k_x$-dependent transmission of the form $T(k_x) = (1+F(k_x)\sin ^2[\theta(k_x)])^{-1} $.  
% $T(k_x) = \frac{1}{1+F(k_x)\sin ^2[\theta(k_x)]} $.$ 
Here $F= 4r/(1-r)^2$ is a finesse coefficient, depending on the probability $r$ of internal reflection at the interfaces. The angle $\theta$ is the semiclassical phase accumulated along a path between the two interfaces.
This approach can be generalized to account for the presence of the tip, which is generally described by a Lorentzian shape (details are provided in the Supporting Information). The potential generated by the tip at a generic point of coordinates $(x,y)$, with tip located in $(x_{\mathrm{tip}},y_{\mathrm{tip}})$, can be written as
\begin{equation}
V_{\mathrm{tip}}(x,y) \simeq \langle V_{\mathrm{tip}} \rangle + v_{\mathrm{tip}}(x,y),
\label{eq:Vtip_2comp}
\end{equation}
where $\langle \dots \rangle$ denotes the spatial average of the Lorentzian profile over the junction, and $v_{\mathrm{tip}}$ represents a local contribution of Lorentzian shape, but of much smaller amplitude than the nominal value of $V_{\mathrm{tip}}$. The term $\langle V_{\mathrm{tip}} \rangle$ is responsible for a global renormalization of the Fermi momentum, which becomes the effective quantity $k_{\text{F,eff}}$ and manifests as a shift of the effective CNP, as shown in Fig.~\ref{Fig2}. The effect of the Lorentzian perturbation $v_{\mathrm{tip}}(x,y)$, by contrast, is local: it deforms the semiclassical phase $\theta$ into $\theta+\delta\theta(x-x_{\mathrm{tip}})$, where $\delta\theta$ depends on the impact parameter $x-x_{\mathrm{tip}}$ of the electrons relative to the tip center. This is, in all respects, a coherent \textit{lensing effect} \cite{BornWolf1999}, in which the tip actively modifies the quantum phase accumulated along each FP path between the two interfaces. 

Within the semiclassical approximation \cite{Shytov2008} for the transmission probability at the interfaces, one gets $r=1-\exp\left(-2 k_x^2/k_{\text{F,eff}}^2\right)$, resulting in an effective finesse $F_{\text{eff}}(k_x, k_{F, \text{eff}})$.
The expression for the conductance thus becomes (see Supporting Information for more details)
\begin{equation}
   G =\frac{4e^2}{ h} \int \frac{ dk_x}{2\pi}\int_{-W/2}^{W/2} dx \,\frac{1}{1+F_{\text{eff}}(k_x,k_{\text{F,eff}})\sin^2 [k_{\text{F,eff}}L_c+\delta\theta(x-x_{\mathrm{tip}})]},
   \label{eq1}
\end{equation}
where the explicit expression for $\delta \theta$ reads
\begin{equation}
    \delta\theta(x-x_{\mathrm{tip}})= 2 (\eta V_{\mathrm{tip}}) \,k_{\text{F,eff}}\frac{R^2}{\sqrt{R^2+(x-x_{\mathrm{tip}})^2}}\arctan\left(\frac{L}{2\sqrt{R^2+(x-x_{\mathrm{tip}})^2}}\right),
    \label{deltatheta}
\end{equation}
with $\eta V_{\mathrm{tip}}$ the dimensionless fraction of Fermi energy induced by the perturbation $v_{\mathrm{tip}}$ and $R$ the Lorentzian radius of the tip potential. This  coherent phase shift gives rise to a complex oscillation pattern that closely reproduces what is experimentally observed in conductance line cuts as a function of $V_{\mathrm{tip}}$ at fixed $k_{\text{F,eff}}$. This is shown in Fig.~\ref{Fig2}g for three different values of $k_{\text{F,eff}}$, indicated by the vertical colored arrows in Fig.~\ref{Fig2}d. An excellent agreement is obtained between the experimental curves (dashed lines) and the theoretical ones (solid lines, considering the tip located at junction center). The main harmonic of the oscillations arises from the single-mode contribution at fixed $x$, whose transmission amplitude follows a form of the type $ [1+F\sin^2(a+bV_{\mathrm{tip}})]^{-1} $. We note that finer oscillations are also superimposed on the main harmonic component. These additional oscillations and the deformation of the simple single-mode curve are due to the intricate integral superposition of all transmissions for trajectories originating from different $x$ values. Moreover, analyzing the dependence of the conductance oscillations on the tip radius, the main harmonic increases systematically as the Lorentzian radius decreases, which supports the lensing interpretation of the incident electron beam (see Supporting Information). 

This analytical model not only is able to explain the nature of the observed oscillations, but also demonstrates that the SGM tip departs from the role of a passive probe, effectively becoming an active agent for the coherent control of the quantum phase of electrons in graphene.

\begin{figure}[!tbh] % Fig.3, caption 162 words
	\centering
	\includegraphics[width=\linewidth]{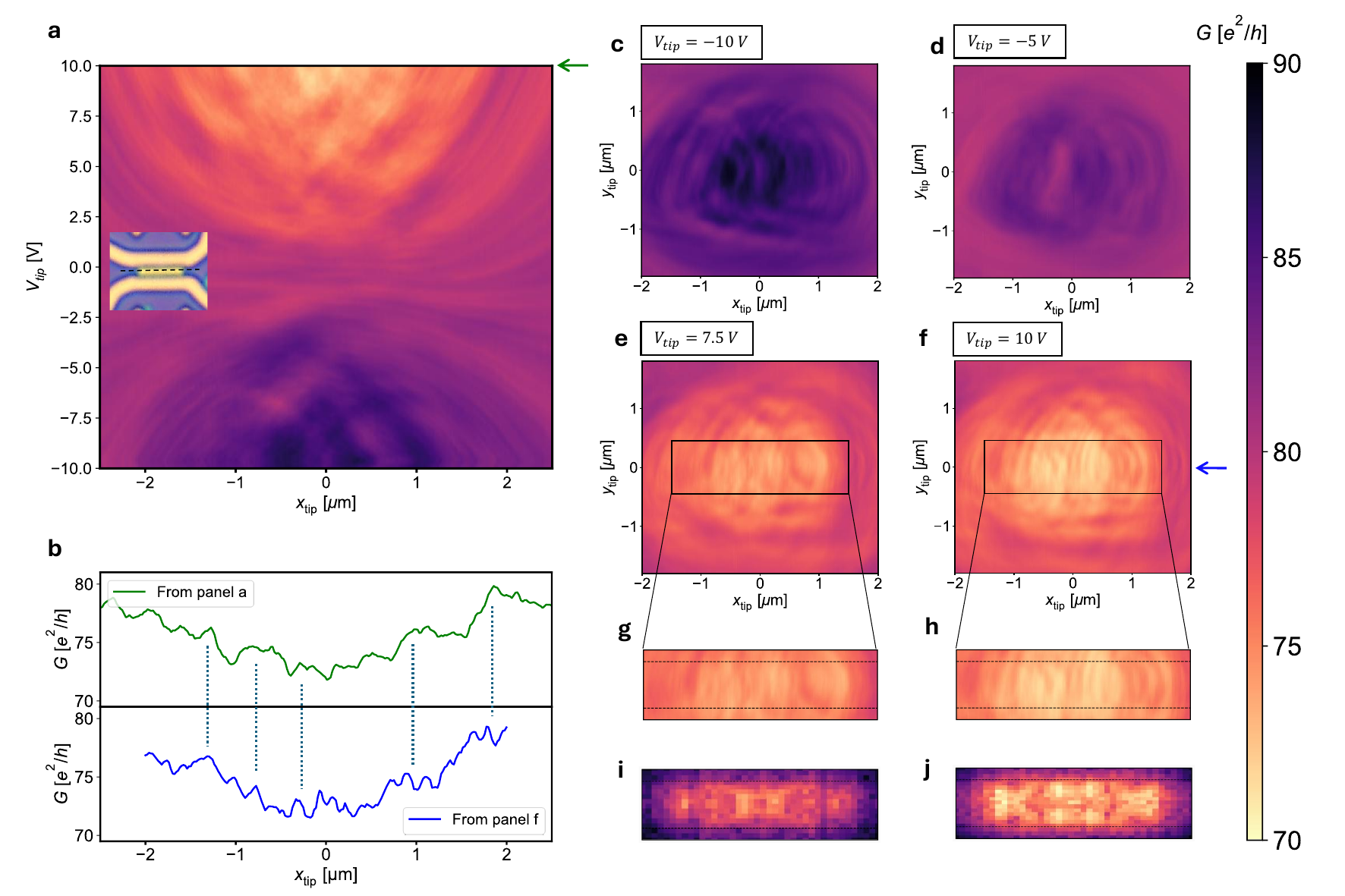}
	\caption{(\label{Fig3}\textbf{a}) Map of junction conductance $G$ as a function of $V_{\mathrm{tip}}$ and $x_{\mathrm{\mathrm{tip}}}$. Tip is scanned in the $x_{\mathrm{\mathrm{tip}}}$-direction along a line at $y_{\mathrm{\mathrm{tip}}}=0$, as indicated in the inset. $V_{\mathrm{BG}}=-20$~V. (\textbf{b}) Comparison of line cuts of $G$ vs. $x_{\mathrm{\mathrm{tip}}}$. Green line is a horizontal line cut extracted from panel \textbf{a} at fixed $V_{\mathrm{tip}}=10$~V (green arrow), blue line is a horizontal line cut extracted from the SGM map of Fig.~\ref{Fig3}f at fixed $y_{\mathrm{\mathrm{tip}}}=0$ (blue arrow). The dashed guiding lines highlight that the curves share the same oscillatory pattern. (\textbf{c}-\textbf{f}) SGM maps of the junction conductance $G$ in the hole doping regime ($V_{\mathrm{BG}}=-20$~V), for different values of $V_{\mathrm{tip}}$. (\textbf{g}-\textbf{h}) Zoom-in of the area in black rectangles in panels \textbf{e}-\textbf{f}. Dashed black lines outline the junction geometry. (\textbf{i}) Simulation of the SGM data in \textbf{g}, for $V_{\mathrm{tip}}=7.5$~V. (\textbf{j}) Simulation of the SGM data in \textbf{h}, for $V_{\mathrm{tip}}=10$~V. The same color scale is used for experimental data (panels \textbf{a} and \textbf{c}-\textbf{h}) and simulations (panels \textbf{i} and \textbf{j}).
    }
\end{figure}

To further explore the tip lensing effect, and eventually connect the observed modulation of FP interference under the combined effect of $V_{\mathrm{BG}}$ and $V_{\mathrm{tip}}$ (shown in Fig.~\ref{Fig2}) to the spatial maps analyzed in Fig.~\ref{Fig1}, we next study the conductance as a function of $V_{\mathrm{tip}}$ and $x_{\mathrm{\mathrm{tip}}}$. This intermediate step fixes one of the voltage parameters of Fig.~\ref{Fig2} ($V_{\mathrm{BG}}=-20$~V) and one of the spatial coordinates of Fig.~\ref{Fig1} ($y_{\mathrm{\mathrm{tip}}}=0$). The resulting map is shown in Fig.~\ref{Fig3}a. Also in this case we observe a global background reduction (enhancement) of $G$ with positive (negative) $V_{\mathrm{tip}}$, arising from the effective shift of the CNP induced by the $\langle V_{\mathrm{tip}}\rangle$ term in Eq.~\eqref{eq:Vtip_2comp}. Along $x_{\mathrm{\mathrm{tip}}}$, the background modulation becomes less pronounced as the tip moves toward the edges of the junction and $\langle V_{\mathrm{tip}}\rangle$ decays. On top of the background, an intricate pattern evolves as a function of both $x_{\mathrm{tip}}$ and $V_{\mathrm{tip}}$, as a result of tip-induced manipulation of FP interference. Indeed, varying $V_{\mathrm{tip}}$ not only modifies $k_{\text{F,eff}}$, but also changes the local term $v_{\mathrm{tip}}$ that enters the lensing phase $\delta\theta$ and ultimately determines the measured conductance (see Eqs.~\eqref{eq1},\eqref{deltatheta}). Similarly, displacing the tip along $x_{\mathrm{\mathrm{tip}}}$ simultaneously modifies $k_{\text{F,eff}}$ and $\delta\theta$. The interplay between these two effects gives rise to the pattern shown in Fig.~\ref{Fig3}a, and demonstrates the broad parameter space available for coherent control of electron motion using the SGM technique. In Fig.~\ref{Fig3}b, we compare the line-cut at fixed $V_{\mathrm{tip}}=10$~V extracted from Fig.~\ref{Fig3}a (green horizontal arrow), with the corresponding line at $y_{\mathrm{tip}}=0$ extracted from the SGM map shown in Fig.~\ref{Fig3}f (blue horizontal arrow). These curves share the same oscillatory behavior, demonstrating that the fringe pattern observed in SGM maps is ascribable to tip-tuned FP interference.

Next, we investigate the evolution of the interference fringes at different $V_{\mathrm{tip}}$ values. SGM maps of $G$ for $V_{\mathrm{tip}}$ ranging from $-10$~V to $+10$~V are shown in Fig.~\ref{Fig3}c to Fig.~\ref{Fig3}f. In all maps, fringe interference patterns are observed, superimposed on a background enhancement or reduction of $G$, depending on the sign of $V_{\mathrm{tip}}$.
The fringe patterns are qualitatively similar, but the specific spatial evolution of the fringes varies with $V_{\mathrm{tip}}$.
The formation of circular fringes can be explained by considering the effect of the spatially averaged term $\langle V_{\mathrm{tip}}\rangle$, which is a function of the tip position, $\langle V_{\mathrm{tip}}\rangle = \langle V_{\mathrm{tip}}(x_{\mathrm{tip}},y_{\mathrm{tip}})\rangle$. Since these coordinates correspond to the pixel coordinates of the experimental maps, each pixel is associated with a different renormalization of the Fermi energy. Equipotential lines of the $\langle V_{\mathrm{tip}}(x_{\mathrm{tip}},y_{\mathrm{tip}})\rangle$ term have a circular shape, which means that along such lines the renormalization of the Fermi momentum is the same. This gives rise to the circular shape of conductance fringes observed experimentally in Fig.~\ref{Fig3}c to Fig.~\ref{Fig3}f and Figs.~\ref{Fig1}c,e. 
However, the conductance is crucially altered by the local lensing effect, which modifies the interference conditions and therefore makes each specific pattern measured by SGM deviate from the equipotential curves of $\langle V_{\mathrm{tip}}\rangle$.
To further verify the lensing effect introduced by the tip, we performed numerical simulations of the SGM maps in Fig.~\ref{Fig3}g ($V_{\mathrm{tip}}=7.5$~V) and Fig.~\ref{Fig3}h ($V_{\mathrm{tip}}=10$~V), shown respectively in Fig.~\ref{Fig3}i and Fig.~\ref{Fig3}j.
These simulations are based on a coarse grained model on a discretized lattice \cite{Lombardi2025,villani_2026_SGM,Liu2014}, with the action of the tip modeled by an effective Lorentzian-shaped potential $ V_{\mathrm{tip}}(\textbf{r}) = V_{\mathrm{tip}} / (1+|\textbf{r}-\textbf{r}_{\mathrm{tip}}|^2/R^2)$ \cite{Szafran2011}. 
Details are provided in the Supporting Information. Besides the global background modulation of $G$, the numerical simulations effectively reproduce the fringe pattern observed in the experimental maps. This allows us to exclude other non-tip-induced effects as the origin of our experimental observations.

Strikingly, the lensing mechanism determines also the pattern observed in Fig.~\ref{Fig1}c for the switching current when the Josephson junction operates in the superconducting regime. In fact, when the junction supports a supercurrent, Andreev bound states form between the Niobium leads. By resonating at the specific superconducting phase difference across the junction, they enable dissipation-less charge transfer. As discussed in \cite{BenShalom2016}, if the electron and hole trajectories cannot form closed loops, phase coherence is lost, and the corresponding Andreev bound state is destroyed, reducing the maximum sustainable supercurrent, i.e., the switching current. The coherent lensing can precisely reproduce this effect. Namely, the tip modifies the phase of both the traveling electrons and holes, and if the superconducting phase difference cannot compensate sufficiently, the tip drives the corresponding Andreev bound state off resonance. This tip-induced off-resonance condition is precisely the same as in FP trajectory lensing, the only difference being that in the normal state a single electron bounces within the cavity. This correspondence shows that Scanning Gate Microscopy can actively control the quantum phase in a wide range of mesoscopic phenomena, spanning from coherent electronic transport to superconductivity.

\newpage
\beginsupplement

\begin{center}
	\section*{Supporting Information}
\end{center}

\section{Fabrication methods}
\label{SI_sec:methods}

The fabrication of graphene Josephson junctions follows the procedure described in Refs.~\cite{Villani2025,villani2026diode_arxiv,villani_2026_SGM}. We employ the dry pick-up method \cite{Wang2013,Purdie2018} to assemble the hBN-graphene-hBN heterostructure. The heterostructure comprises two $30$~nm-thick hBN flakes and is released on a \ce{Si/SiO2} substrate that acts as a back-gate. We employ graphene crystals grown by chemical vapor deposition (CVD) \cite{Miseikis2015,Miseikis2017}. Prior to patterning, the heterostructure is analyzed via Raman spectroscopy and atomic force microscopy (AFM), to identify flat areas with minimal nanoscale strain variations (evaluated from the width of the 2D Raman peak) \cite{Pezzini2020} for device patterning. 

A first electron-beam lithography (EBL) patterning step defines a mask used for etching of the heterostructure and deposition of Niobium (Nb) contacts through magnetron sputtering. This step leads to self-aligned one-dimensional contacts \cite{BenShalom2016}, whose height matches the thickness of the stack, $60$~nm. A second EBL step is used to define the device mesa through a second etching step.

Navigation markers comprising arrows and alphanumeric markers are co-patterned during the first EBL step. As a result, they have the same height as the Nb contacts and the heterostructure, preventing interference during SGM scans. Markers are necessary to navigate on the sample when the system is at $T=300$~mK, because no imaging tools are available, other than topography maps. The markers pitch is optimized in such a way that at least one marker will always be present in the $8\times 8\;\mu$m$^2$ scanning area at $T=300$~mK.

\section{Additional SGM maps of $I_{sw}$}
\label{SI_sec:add_data_Isw}

\begin{figure}[!htb]
	\centering
	\includegraphics[width=0.7\linewidth]{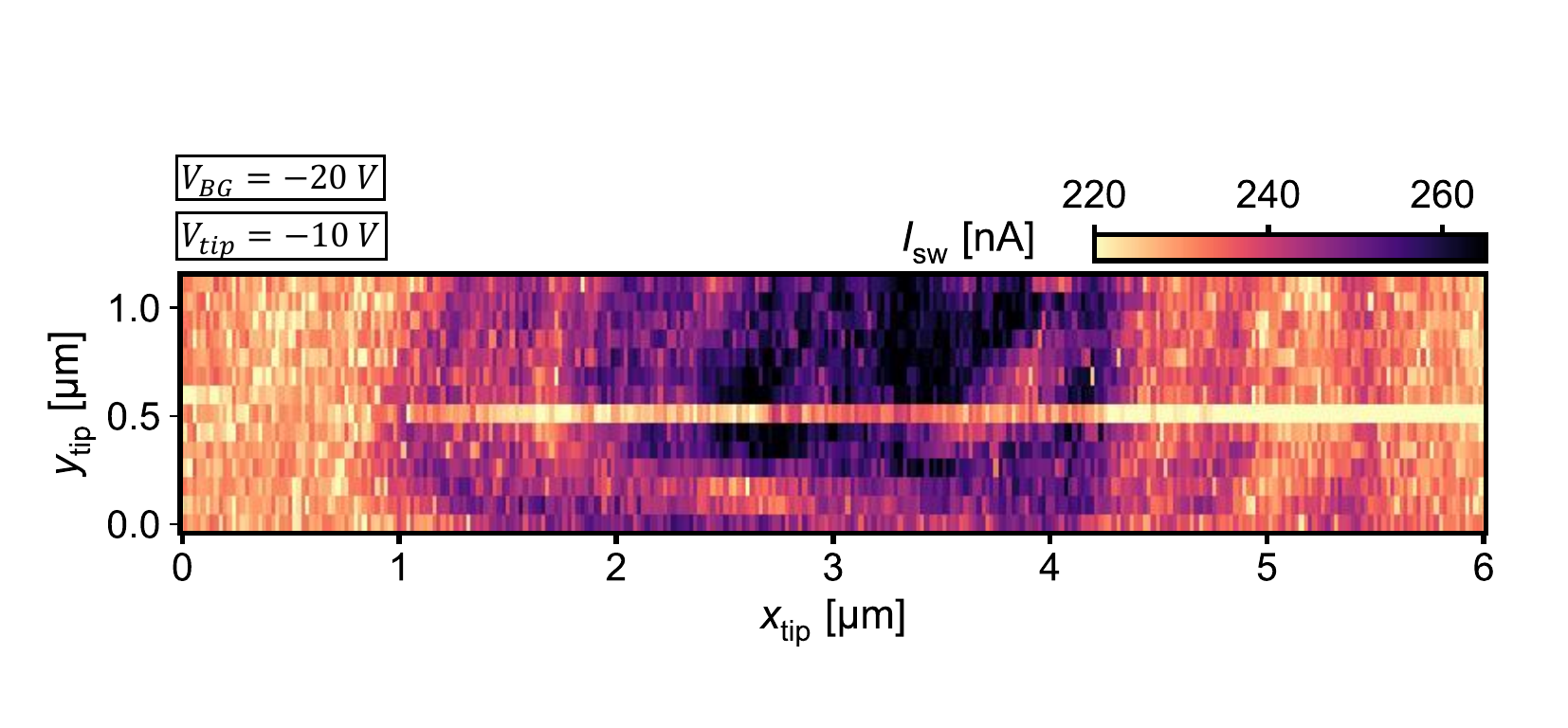}
	\caption{\label{fig_add_Isw_maps}SGM map of switching current $I_{sw}$ as a function of the SGM tip coordinates. Tip bias $V_{tip}=-10$~V, back-gate voltage $V_{BG}=-20$~V, $\Delta V_{BG}=-13.7$~V (p-type doping regime).}
\end{figure}

In Fig.~\ref{fig_add_Isw_maps} we present an additional map of the switching current $I_{sw}$ as a function of the tip position $(x_{\mathrm{\mathrm{tip}}},y_{\mathrm{\mathrm{tip}}})$, at the same back-gate voltage value as in the main text, $V_{BG}=-20$~V, but opposite tip bias, $V_{\mathrm{tip}}=-10$~V. As expected for a negative tip bias, we observe an enhancement of the switching current as the tip is moved toward the junction center. Furthermore, similarly to Fig.~1c of the main text, circular fringes of approximately constant switching current are present, superimposed to the global background modulation.

\section{Correlation of lock-in maps with DC maps}
\label{SI_sec:corr_Glockin_DC}

In the main text we show that the switching current $I_{sw}$ and the conductance $G$ obtained from $V$-$I_{bias}$ curves share the same spatial modulation and the same fringe pattern, as further exemplified by the linecut comparison in Fig.~1f. 

Here we demonstrate that measurements of the junction conductance $G$ performed either by extracting this value from DC bias measurements (i.e., from $V$-$I_{\mathrm{bias}}$ curves) or by measuring it with a lock-in amplifier give consistent results, and can therefore be considered equivalent. For clarity, in the following we label as $G_{DC}$ the conductance extracted from $V$-$I_{bias}$ curves as in Fig.~1e of the main text, $G_{LI}$ the conductance measured with a lock-in amplifier (as in Fig.~3c to Fig.~3f of the main text).

\begin{figure}[!htb]
	\centering
	\includegraphics[width=\linewidth]{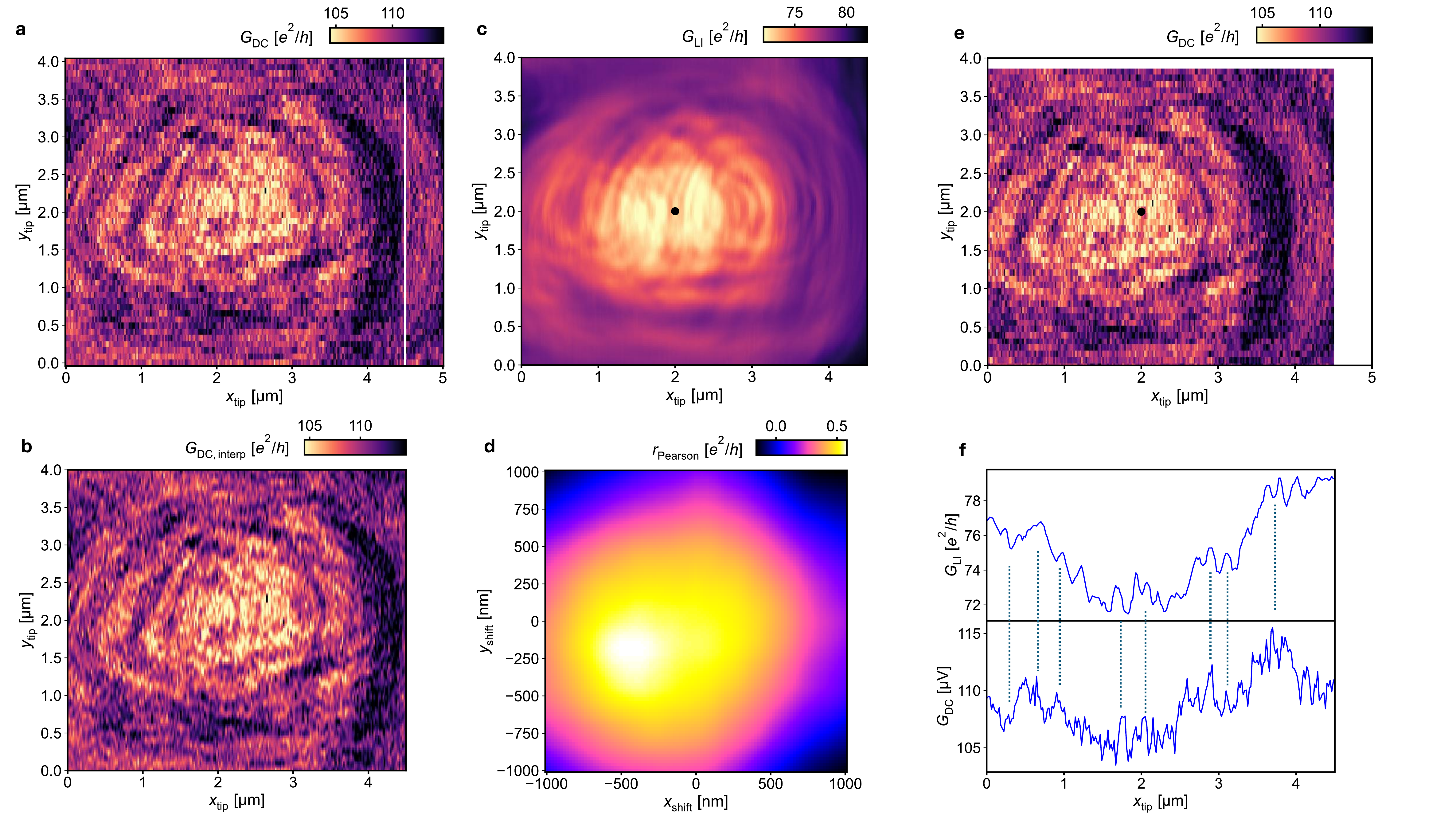}
	\caption{(\textbf{a}) SGM map of junction conductance $G_{DC}$ obtained by fitting the dissipative branch of $V$-$I_{bias}$ curves (from the same dataset as in Fig.~1e of the main text). $V_{BG}=-20$~V, $V_{tip}=10$~V. (\textbf{b}) Interpolated SGM map of junction conductance $G_{DC}$, such that pixels have size $20\times 20$~nm. (\textbf{c}) SGM map of junction conductance $G_{LI}$ measured with a lockin amplifier (from the same dataset as in Fig.~3f of the main text). (\textbf{d}) Pearson correlation coefficient between $G_{LI}$ and $G_{DC}$ maps as a function of the relative spatial shift between the two maps. The shift is the translation of the $G_{DC}$ map with respect to the $G_{LI}$ map. (\textbf{e}) Map of $G_{DC}$ shifted by the optimal shift coordinates extracted from panel \textbf{d}. (\textbf{f}) Comparison of linecuts of $G_{LI}$ (upper panel) and $G_{DC}$ (lower panel) at $y_{tip}=2\;\mu$m, showing that $G_{LI}$ and $G_{DC}$ share the same oscillatory pattern. The $G_{DC}$ linecut is obtained as the average of three lines around $y_{\mathrm{\mathrm{tip}}}=2\;\mu$m. \label{fig_corr_G}}
\end{figure}

To directly compare $G_{DC}$ with $G_{LI}$ an additional alignment step between the maps is necessary. In fact, while $I_{sw}$ and $G_{DC}$ are automatically aligned because the two quantities are extracted from the same dataset, this is not necessarily true for maps from different datasets, as it is the case $G_{LI}$ and $G_{DC}$. Every thermal cycle and tip approach routine can lead to a small drift of the scan area with respect to previous SGM maps, making the superposition between different datasets not exact. As a figure of merit to evaluate the optimal alignment between the maps, we calculate the Pearson correlation coefficient using the \texttt{scipy.stats.pearsonr} Python function, and we search for the shift that maximizes the correlation coefficient. For two-dimensional maps, this parameter can be calculated by first unraveling each two-dimensional map into a one-dimensional ordered array.

The $G_{DC}$ and $G_{LI}$ maps have different scan size and resolution: the $G_{DC}$ map has pixels of $20\times 80$~nm over an area of $5\times 4\;\mu$m$^2$ along $x_{\mathrm{tip}}$ and $y_{\mathrm{tip}}$ directions, while the $G_{LI}$ map has pixels of $20\times 20$~nm, over an area of $4.5\times 4\;\mu$m$^2$. As a first step, we interpolate the $G_{DC}$ map in order to match the $G_{LI}$ map resolution, and crop it such that the range along the $x_{\mathrm{tip}}$-axis matches that of the smaller $G_{LI}$ map. The original $G_{DC}$ map is shown in Fig.~\ref{fig_corr_G}a, the interpolated cropped map (corresponding to the area in Fig.~\ref{fig_corr_G}a limited by the vertical white line) is shown in Fig.~\ref{fig_corr_G}b. The correlation analysis is done in the $(0,4.5)\;\mu$m range along $x_{\mathrm{tip}}$ (and full range along $y_{\mathrm{tip}}$). In Fig.~\ref{fig_corr_G}c we report the original $G_{LI}$ map, that has the same spatial extension of the $G_{DC}$ map in  Fig.~\ref{fig_corr_G}b. We next iterate the calculation of the Pearson correlation coefficient for varying spatial shifts, searching for the $(x_{shift},y_{shift})$ coordinates that maximize the correlation coefficient. The map of the correlation coefficient as a function of $(x_{shift},y_{shift})$ is shown in Fig.~\ref{fig_corr_G}d. The $(x_{shift},y_{shift})$ coordinates indicate the spatial shift of the $G_{DC}$ map with respect to the $G_{LI}$ map. The optimal shift coordinates are $(x_{shift},y_{shift})=(-500,-180)$~nm. The resulting optimally shifted $G_{DC}$ map is shown in Fig.~\ref{fig_corr_G}e.

In Fig.~\ref{fig_corr_G}f we compare line cuts taken at $y_{tip}=2\;\mu$m. Analogously to line cuts of $I_{sw}$ and $G$ shown in Fig.~1f of the main text, local maxima and minima in $G_{LI}$ and $G_{DC}$ are aligned, as exemplified by the guiding black dashed lines, and demonstrate that it is equivalent to measure the junction conductance either with a lock-in amplifier or by extracting the conductance from $V$-$I_{bias}$ curves. This comparison also underlines the exceptional reproducibility and stability (both spatial and temporal) of the observed fringe patterns.

We finally identify the $(2,2)\;\mu$m as the coordinates of the junction center (black dots in Fig.~\ref{fig_corr_G}c and Fig.~\ref{fig_corr_G}e). SGM spatial maps in the main text (Figs.~1c,e and Figs.~3c to Fig.~3f) are plotted such that the center is shifted to $(0,0)$ coordinates, and cropped so that they are symmetric around this point. As a result, the $I_{sw}$ and $G_{DC}$ maps of Figs.~1c,e extend in a $5\times 3.6\;\mu$m$^2$, while $G_{LI}$ maps shown in Fig.~3c to Fig.~3f extend over a $4\times 3.6\;\mu$m$^2$ area.

\section{Polynomial background removal for $\Delta G$}
\label{SI_sec:poly_bg_for_DeltaG}

\begin{figure}[!htb]
	\centering
	\includegraphics[width=0.7\linewidth]{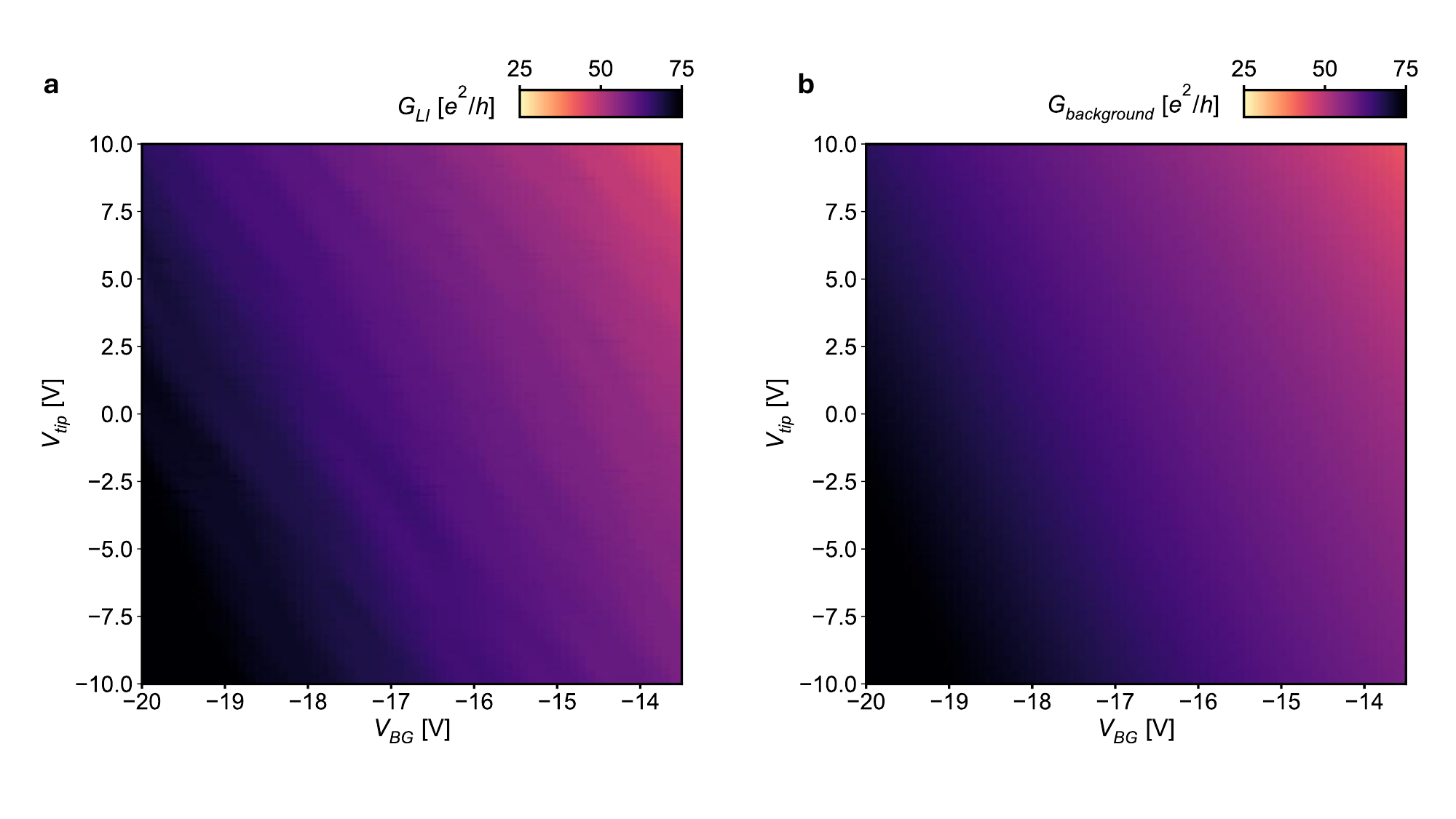}
	\caption{(\textbf{a}) Zoom of Fig.~2b of the main text, in the $V_{BG}=(-20,-13)$~V range. (\textbf{b}) Polynomial background, obtained by performing a linear least-squares fit to the data in panel \textbf{a}. \label{fig_background}}
\end{figure}

To isolate Fabry-Pérot oscillations in $\Delta G$ shown in Fig.~2c of the main text, a polynomial background is removed. This is obtained by fitting a polynomial surface to the raw conductance map $G(V_{\text{BG}}, V_{\text{tip}})$ shown in Fig.~\ref{fig_background}a using a linear least-squares fit. The resulting smooth background $G_{\text{background}}$ (Fig.~\ref{fig_background}b) is then subtracted from the measured data to yield the residual conductance oscillations, $\Delta G = G - G_{\text{background}}$.

\section{Analytical model for the lensing effect}

In this section we show the details of the analytical model that demonstrates the lensing effect. We extend the model described in \cite{Shytov2008}, where authors consider translation invariance along the interfaces. Since the tip breaks this symmetry, we here performed a spatially resolved scattering analysis. Defining $x$ as the coordinate parallel to the cavity interfaces, one can consider scattering modes with well-defined $x$ (see Fig. \ref{fig_corr_G}, Fig. \ref{fig:due_pannelli}). For the Heisenberg uncertainty principle, these modes consist of a quantum superposition of momenta $k_x$. Since the interfaces, through the Klein collimation \cite{Shytov2008,Young2009}, act as a filter for the quasi-normal modes $k_x \ll k_F$, and if the tip potential varies slowly with respect to $k_F$ in the $x$-direction, one can assume that the non-filtered part of the superposition in $k_x$ is \textit{locally} conserved, where locally means that the conservation is true at a fixed $x$. The Dirac equation for the sublattice spinor $\psi$ of a scattering mode with well-defined $x$ then reads
\begin{equation} i\partial_y \psi = \left[(\mu_p+V_{\text{tip}}(x,y))\sigma_z-ik_x\sigma_x\right]\psi, \end{equation}
where $\mu_p$ is the chemical potential of the $p-$cavity, and $\sigma_i$ are the Pauli matrices. Here, without loss of generality, we set the Fermi energy to zero. From the Dirac equation follows $k_y(x,y)=\sqrt{(\mu_p+V_{\text{tip}}(x,y))^2-k_x^2}$. The scattering amplitude $A_{\textbf{r} \,\rightarrow \textbf{r}_f}$ from $\textbf{r}$ (at the first encountered interface) to $\textbf{r}_f$ (on the opposite interface) can be written in a path-integral fashion
\begin{equation} A_{\textbf{r} \,\rightarrow \textbf{r}_f} \propto \int\frac{dk_x}{2\pi}e^{ik_x(x_f-x)}\sum_{\text{FP paths}}e^{i S[x,\,k_x]/\hbar}, \end{equation}
where the sum is over the Fabry-Pérot paths and $S$ is the action \cite{Feynman}.
 This sum is a geometric series, and so the amplitude results in 
\begin{equation} A_{\textbf{r} \,\rightarrow \textbf{r}_f} \propto \int\frac{dk_x}{2\pi}e^{ik_x(x_f-x)}\frac{t(k_x,\mu_p) e^{i \theta_{\text{WKB}}(x,k_x)}}{1-r(k_x,\mu_p)e^{2i\theta_{\text{WKB}}(x,k_x)}}, \end{equation}
where $t=1-r$ is the transmission probability across the interface, and
\begin{equation} \theta_{\text{WKB}}(x,k_x) = \int_{-L_c}^{L_c}k_y(x,y,k_x)dy. \end{equation}
Performing the approximation for $V_{\text{tip}}(x,y) \simeq \langle V_{\text{tip}}\rangle + v_{\text{tip}}(x,y)$ and a perturbative expansion in $\delta V_{\text{tip}}/(\mu_p+\langle V_{\text{tip}}\rangle)$, the WKB angle \cite{WKB} reduces to $\theta_{\text{WKB}}= \theta_{eff}+\delta\theta(x,k_x)$. The conductance is
\begin{equation} G\propto \int_{-W/2}^{W/2}dx\,dx_f\,\left|A_{\textbf{r} \,\rightarrow \textbf{r}_f}\right|^2, \end{equation}
and this leads to 
\begin{equation}G =\frac{4e^2}{ h} \int \frac{ dk_x}{2\pi}\int_{-W/2}^{W/2} dx \,\frac{1}{1+F_{\text{eff}}(k_x,k_{F,\text{eff}})\sin^2 [k_{F,\text{eff}}L_c+\delta\theta(x-x_{\mathrm{tip}})]}.\end{equation}
The explicit expression for $\delta\theta(x-x_{\mathrm{tip}})|_{y_{\text{tip=0}}}$ is

\begin{equation}
    \delta\theta(x-x_{\mathrm{tip}})= 2 (\eta V_{\mathrm{tip}}) \,k_{F,\text{eff}}\frac{R^2}{\sqrt{R^2+(x-x_{\mathrm{tip}})^2}}\arctan\left(\frac{L}{2\sqrt{R^2+(x-x_{\mathrm{tip}})^2}}\right).
    \label{deltatheta_SI}
\end{equation}
In Fig.~2g of the main text we have shown how this formula predicts the conductance oscillations at a fixed effective Fermi momentum. In addition to the main harmonic, higher frequency oscillations are observed, which arise from the $x$-integral over the transmission amplitudes $T(x,k_x)$ at a fixed $k_x$. Indeed, the $x-$dependence of the function $T(x,k_x)= (1+F_{\text{eff}}(k_x,k_{F,\text{eff}})\sin^2 [k_{F,\text{eff}}L_c+\delta\theta(x-x_{\mathrm{tip}})])^{-1}$ exhibit a spatial pattern that reflects how a single mode feels the local potential of the tip modifying its phase. Integrating over the different patterns for each $k_x$ results in the generation of the higher harmonics in the conductance. Here, $L_c$ represents an effective length over which electrons start to feel the effect of the tip.

\begin{figure}[!htb]
    \centering
    \includegraphics[width=0.5\linewidth]{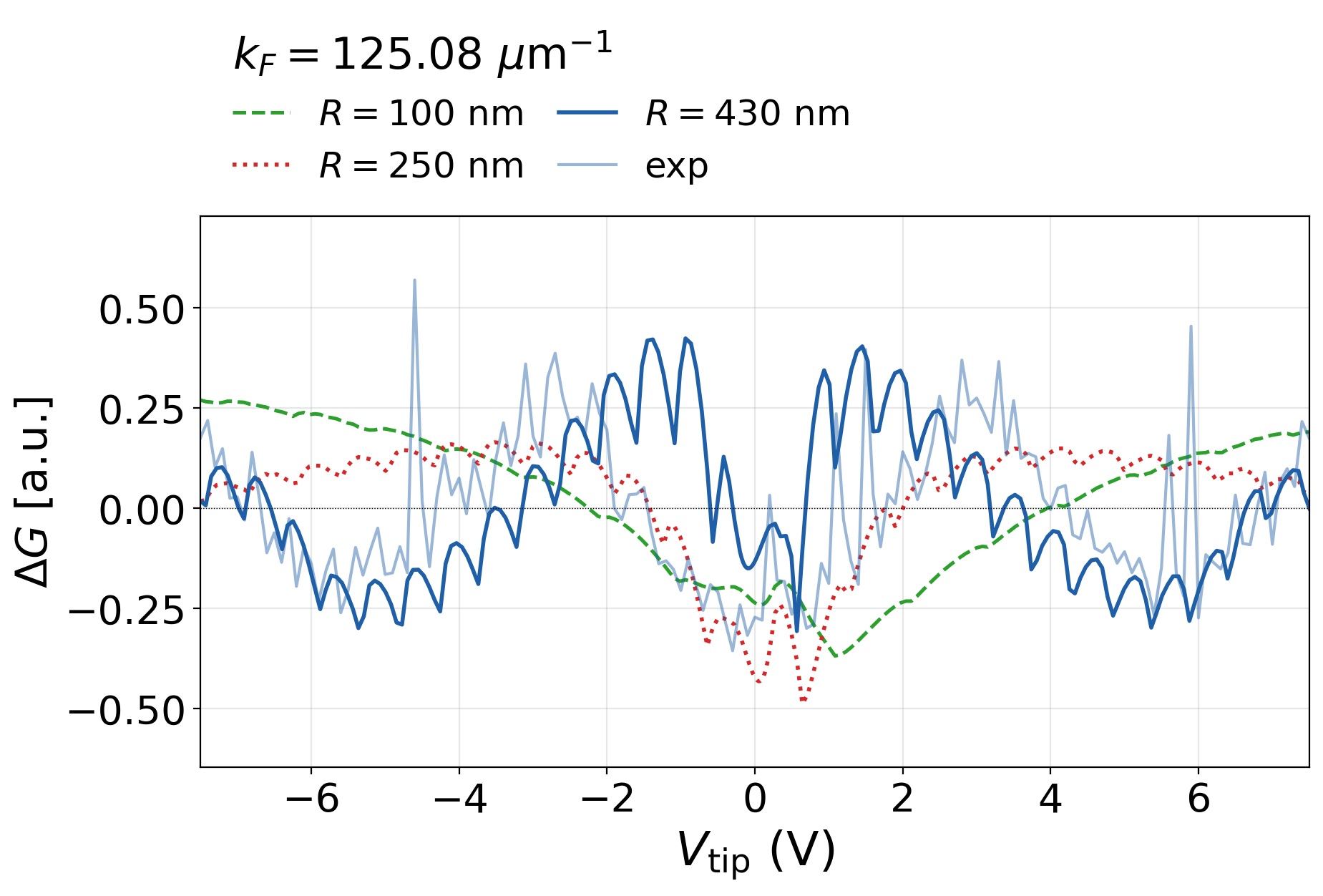}
    \caption{Comparison of conductance oscillations for different tip radius $R$. Lowering the radius means increasing the period of the main harmonic, showing that the oscillations arise from the lensing of a sufficiently large number of incident modes with impact parameter $x-x_{\text{tip}} \lesssim R$. A comparison with the experimental curve is shown (light blue line).}
    \label{S5_1}
\end{figure}

To further demonstrate the role of the tip in the lensing effect, we show how the conductance modulation depends on the Lorentzian radius. In Fig.~\ref{S5_1} we show, as a function of the tip radius $R$, the conductance $G(V_{\text{tip}})$ after the polynomial background removal for a fixed $k_{F,\text{eff}}$. As the radius is lowered, the main harmonic increases its period, reflecting the fact that less scattering modes have an impact parameter $x-x_{\text{tip}} \lesssim R$ that contribute to the main interference effect.

\label{SI_sec:analyt_model}

\section{Details on the numerical model}

\begin{figure}[!ttb]
    \centering
    \begin{subfigure}[b]{0.18\linewidth}
        \centering
        \includegraphics[width=3\linewidth]{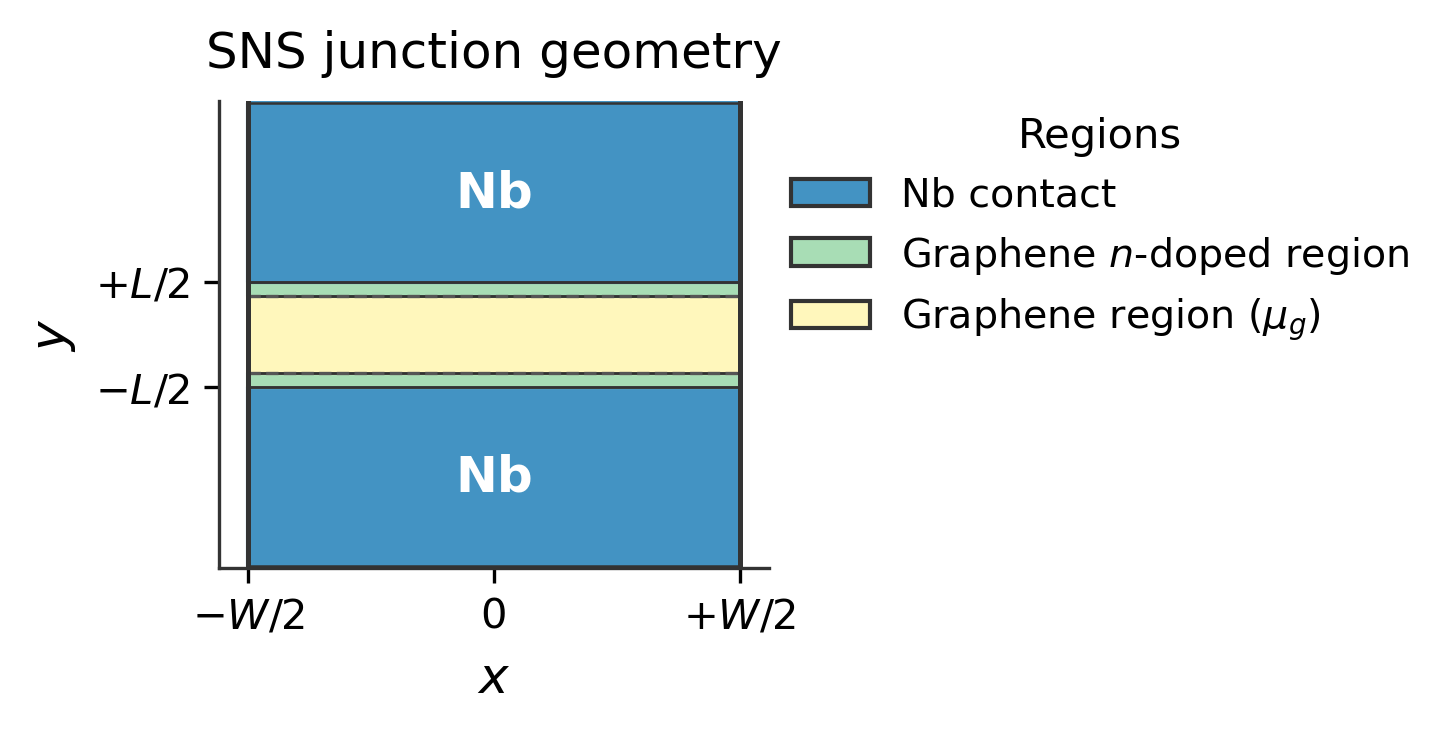}
        \caption{}
        \label{sketchSNS}
    \end{subfigure}
    \hfill
    \begin{subfigure}[b]{0.48\linewidth}
        \centering
        \includegraphics[width=0.8\linewidth]{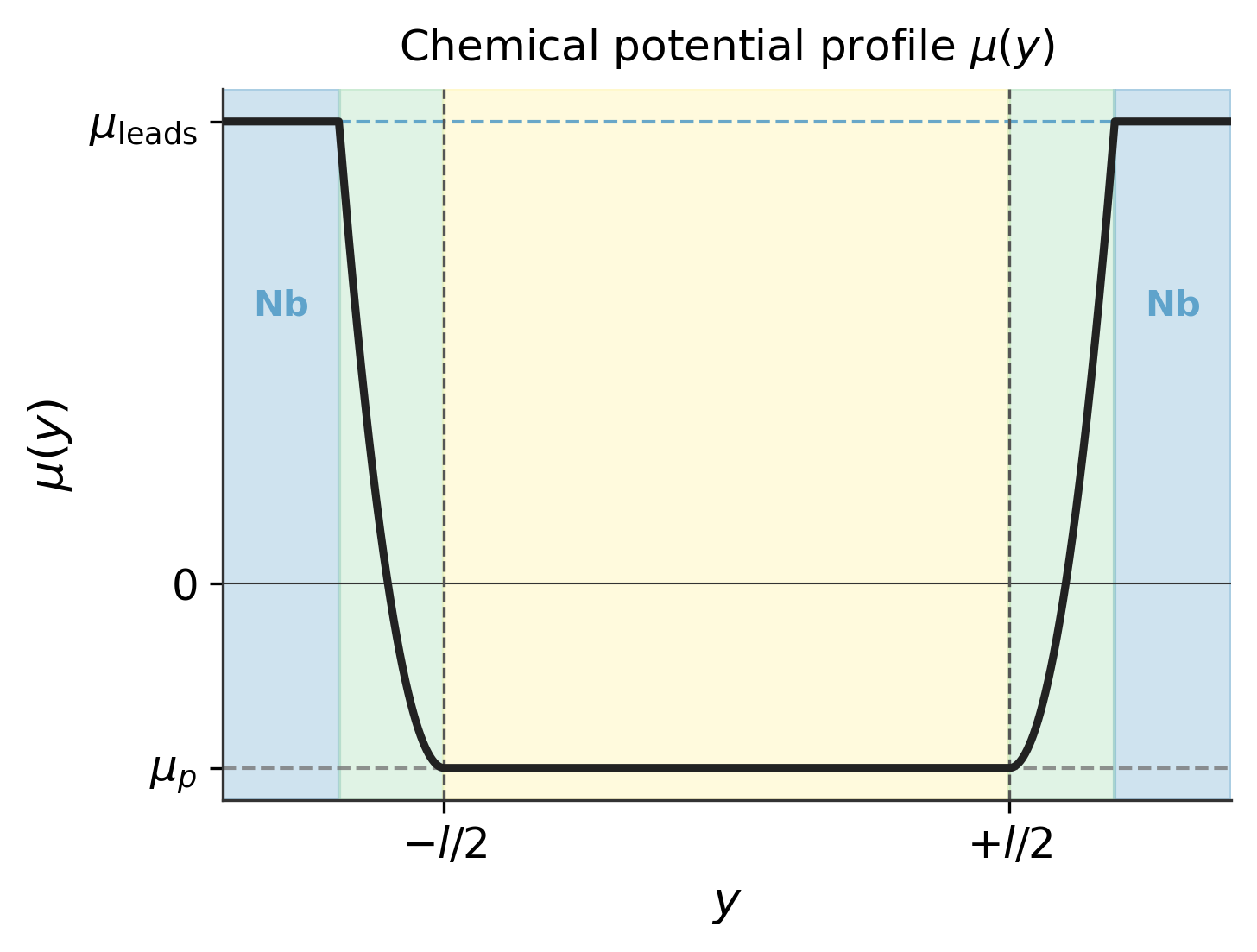}
        \caption{}
        \label{tuo_label_b}
    \end{subfigure}
    \caption{Panel \textbf{a} shows a top-view sketch of the graphene junction.
    Blue regions represent the superconducting contacts with Fermi energy $\mu_{\text{leads}}$.
    Green regions are the $n$-doped parts of graphene with smoothly interpolating
    chemical potential $\mu(y)$. The yellow region is the graphene area with flat
    Fermi energy $\mu_p$, tunable via the backgate voltage.
    Panel \textbf{b} shows explicitly the spatial dependence of $\mu(y)$ along the junction.}
    \label{fig:due_pannelli}
\end{figure}

We describe here the theoretical framework underlying the simulations presented in the main text. We consider a planar junction lying in the $x$-$y$ plane, with $y$ as the longitudinal coordinate parallel to the current flow. In Fig.~\ref{fig:due_pannelli}a we show a sketch of the simulations setup. To avoid possible edge bound states, not present in the experiment, we have chosen armchair edges along $x$ for the model of graphene. Its width is $W \times a$ along $x$ and its length $L \times a$ along $y$, where $a$ is the coarse-grained lattice constant \cite{Lombardi2025,villani_2026_SGM,Liu2014,Liu2026}. The niobium interfaces are located at fixed values of $y$ and extend along $x$. The tight-binding Hamiltonian is defined following Ref.~\cite{Lombardi2025,villani_2026_SGM}. Because of the coarse-graining of $a$, the hopping amplitude $t$ must be renormalized~\cite{Liu2014,Liu2026} so that the graphene Fermi velocity $v_F \propto at$ is kept constant.

To model the Nb metallic contacts, we introduce a spatially modulated chemical potential $\mu = \mu(y)$, with
\[ \mu(y<-L/2\times a)= \mu(y>L/2 \times a) \equiv \mu_{\text{leads}}. \]
To account for the doping induced by the Nb contacts, we define a central region $-l/2\times a < y < l/2 \times a$ with a flat chemical potential $\mu_p$, representing the backgate-tunable Fermi energy of the graphene sheet. At each interface with the leads, $\mu(y)$ interpolates smoothly between $\mu_{\text{leads}}$ and $\mu_p$ over a region of length $m\times a$, with $L=2m+l$. These transition regions correspond to the $n$-doped parts of graphene induced by proximity to the Nb contacts. Fig.~\ref{sketchSNS} shows a sketch of the junction (panel \textbf{a}) and the corresponding $\mu(y)$ profile (panel \textbf{b}).

Disorder is modeled as a sum of Gaussian puddles,
\[ \mu_{dis}(x,y) = \sum_{n=1}^N v_{n}\exp\left(-\frac{(x-x_n)^2+(y-y_n)^2}{\xi^2}\right), \]
where each amplitude $v_n$ is drawn from a Gaussian distribution with standard deviation
$V_{dis}$, each puddle has correlation length $\xi$, and the centers $(x_n,y_n)$ are
uniformly distributed at random over the graphene region. Their number $N$ is set by
requiring $N\pi \xi^2/WL \equiv f$, i.e., a fixed areal coverage fraction $f$.

The SGM tip is modeled as a local modification of the Fermi energy background. Following
Refs.~\cite{Szafran2011,Herbschleb2015,Lombardi2025,villani_2026_SGM}, the tip potential takes the
Lorentzian form
\begin{equation}
\mu_{tip}(x,y) = \frac{\mu_{0,tip}}{1+\dfrac{x^2+y^2}{R^2}},
\label{eq_SI:Lorentzian}
\end{equation}
where $R_{Lor}$ is the effective tip radius. To relate $\mu_p$ and $\mu_{0,tip}$ to the
experimental gate voltages, we note that in graphene the charge density
$n = \mu^2/[\pi(\hbar v_F)^2]$ is a linear function of the gate voltage, which yields
$d\mu_p^2/dV_{bg} = \text{const}$ and hence
\[ \mu_p(V_{bg}) = \alpha\,\mathrm{sgn}(V_{bg})\,\sqrt{|V_{bg}|}, \]
\[ \mu_{0,tip}(V_{tip}) = \beta\,\mathrm{sgn}(V_{tip})\,\sqrt{|V_{tip}|}. \]
These relations fully specify the mapping between the simulation parameters and the
experimental control knobs. The conductance is computed in the linear-response regime
via the Landauer formula
\[ G = \frac{e^2}{h}\,\mathrm{Tr}\,tt^\dagger, \]
where $t$ is the transmission block of the scattering matrix between the two leads.
All simulations are performed using \textsc{Kwant}~\cite{Groth_2014}. The parameter
values are listed in Table~\ref{tab:params}.
\begin{table}[ht]
\centering
\caption{Summary of simulation parameters.}
\begin{tabular}{llll}
\hline
\hline
Parameter & Symbol & Value\\
\hline
Primitive vector length & $a$ & \SI{14}{\nano\meter} \\
Hopping energy & $t$ & \SI{67.5}{\milli\electronvolt} \\
Sample width & $W$ & \SI{3}{\micro\meter}\\
Sample length & $L$ & \SI{600}{\nano\meter} \\
Normal region length & $l$ & \SI{440}{\nano\meter}\\
Interface length & $m$ & \SI{80}{\nano\meter}\\
Fermi energy of leads & $\mu_{\text{leads}}$ & \SI{64}{\milli\electronvolt} \\
Conversion $\mu_p-V_{bg}$ & $\alpha$ & \SI{27}{\milli\electronvolt\times\volt^{-1/2}}\\
Conversion $\mu_{0,\text{tip}}-V_{\text{tip}}$ & $\beta$ & \SI{21.6}{\milli\electronvolt\times\volt^{-1/2}}\\
Tip effective radius & $R$ & \SI{436}{\nano\meter}\\
Standard deviation of disorder strength & $V_{dis}$ & \SI{1.35}{\milli\electronvolt}\\
Disorder correlation length & $\xi$ & \SI{60}{\nano\meter} \\
Disorder filling fraction & $f$ & 10\% \\
\hline
\hline
\end{tabular}
\label{tab:params}
\end{table}

\label{SI_sec:numer_model}

\newpage
\printbibliography

\section*{Acknowledgments}

We acknowledge Simone Traverso for useful discussions, and support from projects PRIN2022 2022-PH852L\-(PE3) TopoFlag and PNRR MUR Project No. PE0000023-NQSTI. K. W. and T. T. acknowledge support from the CREST (JPMJCR24A5), JST and World Premier International Research Center Initiative (WPI), MEXT, Japan.

% \section*{Supporting information}

% A listing of the contents of each file supplied as Supporting Information should be included. For instructions on what should be included in the Supporting Information as well as how to prepare this material for publications, refer to the journal's Instructions for Authors.

% The following files are available free of charge.
% \begin{itemize}
%   \item "Supp\_info\_coh\_elec\_lensing": fabrication methods; additional SGM map of the switching current; correlation between lock-in and DC maps of the junction conductance; polynomial background removal from the conductance map; details on the analytical model for the lensing effect; details of the numerical model.
%   \item "Dataset\_coh\_elec\_lensing": numerical data of Fig.~\ref{Fig1}c to Fig.~\ref{Fig1}f; Fig.~\ref{Fig2}a to Fig.~\ref{Fig2}g; Fig.~\ref{Fig3}a to Fig.~\ref{Fig3}j.
% \end{itemize}

\end{document}